\documentclass[aps,pre,superscriptaddress,groupedaddress,showpacs,twocolumn]{revtex4}
\usepackage{graphicx,epsf}
\usepackage{hyperref}
\usepackage{bbm}
\usepackage{xcolor}
\usepackage{psfrag}
\usepackage{amsmath}
\usepackage{amssymb}
\usepackage{orcidlink}
\usepackage{graphicx,epsf}
\usepackage[english]{babel}
\usepackage[title]{appendix}
  
\begin{document}

\title{Modeling adsorption processes on the core-shell-like polymer structures: star and comb topologies}

\author{Viktoria Blavatska\orcidlink{0000-0001-6158-1636}
}
\affiliation{Institute for Condensed
Matter Physics of the National Academy of Sciences of Ukraine,\\
1 Svientsitskii Str., UA-79011 Lviv, Ukraine}
\email{blavatskav@gmail.com}
\author{Jaroslav Ilnytskyi\orcidlink{0000-0002-1868-5648}
}
\affiliation{Institute for Condensed
Matter Physics of the National Academy of Sciences of Ukraine,\\
1 Svientsitskii Str., UA-79011 Lviv, Ukraine}
\affiliation{Institute of Applied Mathematics and Fundamental Sciences, \\
Lviv Polytechnic National University, 12 S. Bandera Str., UA-79013 Lviv, Ukraine}
\author{Erkki L\"ahderanta
	\orcidlink{0000-0002-1596-2849}
}
\affiliation{Department of Physics, School of Engineering Science,
LUT University, Yliopistonkatu 34, FI-53850 Lappeenranta, Finland}

\begin{abstract}
Coagulation-flocculation of pollutants and chelation of heavy metal ions are two widely used techniques in wastewater purification. Despite the differences between their respective mechanisms and inherent length scales, they bear much similarity on a larger scale, and can both be treated as adsorption of obstacles on a polymer structure. In this regime, their adsorbing efficiency is predominantly affected by conformation statistics of involved polymers, and this approach has been used in our previous studies based on lattice polymer model for a linear polymer adsorbent. There is a strong experimental evidence that branched adsorbents are more efficient than their linear counterparts. In this study we focus on two simplest representatives of the core-shell branched architectures: the star-like (with zero-dimensional point-like core) and comb-like polymers (with one-dimensional rigid core) with various number of branches, $f$, branch lengths, $N$, and branches separations, $S$ (for the case of comb-like structure). The polymers are grown on a lattice using the Monte Carlo simulations with the pruned-enriched Rosenbluth algorithm. 
The quantitative estimates for adsorption capacity $ {\overline  {\langle n_a  \rangle}}$  in terms of adsorbed obstacles per monomer and the average number of bonds ${\overline  {\langle n_{{\rm bond}} \rangle}}$ per adsorbed particle (average adsorption strength) have been evaluated in a wide range of 
parameters $f$, $N$, and $S$. 
Both the case of implicit diffusion of obstacles (with averaging over different arrangements of immobilized obstacles) and explicit diffusion of obstacles (allowing to study dynamics of adsorption process) have been analyzed.   
We found that comb-like polymers display the higher adsorption capacity but lower adsorption strength, comparing to the star-like polymers, and these effects are more pronounced with increasing branches separations $S$. 
Our analysis indicates essential role of bridging between adjacent branches by shared adsorbed particles.

\end{abstract}

\maketitle

\section{Introduction}\label{I}

Flocculation of coagulated impurities (microflocs) \cite{Lee2014, Deng2022, Asgari2023} and chelation of heavy metal ions \cite{Singh2018, Qasem2021} are two widely used processes for wastewater purification \cite{Singh2024}. These two have much in common being based on adsorption of a pollutant on the polymeric structure, but occur on different length scale. Indeed, microflocks are of tens/hundreds of micrometers in size, whereas metal ions, involved in chelation, are of atomic scale. The nature of the polymer-impurity attraction is also different: electrostatic/dispersion forces vs formation of chelation complexes \cite{Butovych2024}, respectively. These differences, however, are much reduced if one considers length scale that is large enough, where a principal role is played by two factors: the statistics of polymeric conformations and the ability of a pollutant to adhere to its monomers. Therefore, one may attempt to consider both processes by the same model paradigm but for appropriately rescaled unit length that is based on the pollutant dimensions. Such approach has been done in our previous studies \cite{Blavatska2024a, Blavatska2024b} by introducing generalized terms of polymer ``adsorbent" and ``obstacles", and combining concepts of the self-avoiding walk and diffusion-limited aggregation on a lattice \cite{deGennes1979, desCloizeaux1982}. These studies covered the case of linear adsorbent.

\begin{figure}[htb]
\begin{center}
\includegraphics[width=9cm]{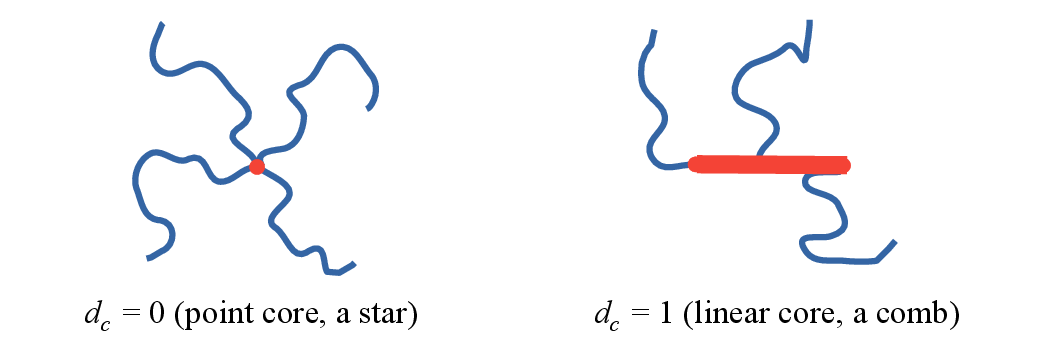}\\
\end{center}
\caption{\label{gener_stars}Two simplest core-shell-like structures of star- and comb-like polymers considered in this study.}
\end{figure}
There is a strong experimental evidence that branched polymeric adsorbents are more efficient than their linear counterparts. The most relevant cases are: $10$-fold increase in the overall stability constant, $K_4$, for the chelation complex between branched polyethylenimine (PEI) and a wide range of metal ions \cite{Kobayashi1987} comparing to linear PEI; triple increase of adsorption capacity for amidoxime functionalized star-like polymers for uranium extraction from seawater, comparing to the linear chains \cite{Yi2024}; an increase in both floc stability and reflocculation ability for highly charged branched cationic polyacrylamide flocculant \cite{Blanco2009}. The other cases to mention are: efficient self-assembled star-like polymeric micelles for chelating Fe(III) \cite{Liu2017}; a membrane of cross-linkable star-like polymer for removal of hexavalent chromium ions from wastewater \cite{Jo2022}; hydrazide-functionalized star-like polymers for efficient recovery and recycling/upcycling of precious metals \cite{Shin2024}, etc. Enhancement of adsorption abilities by branching of a polymeric adsorbent can be attributed to such factors as: the presence of denser regions involving microcages for confining obstacles; the presence of the branching fragments that are naturally suitable for complexation and don't require bending as in the case of a linear chain; reduction of the chains entanglements and, as the result, better accessibility of adsorbing sites, etc. \cite{Mouhamad2023}.

\begin{figure}[htb]
\includegraphics[width=8cm]{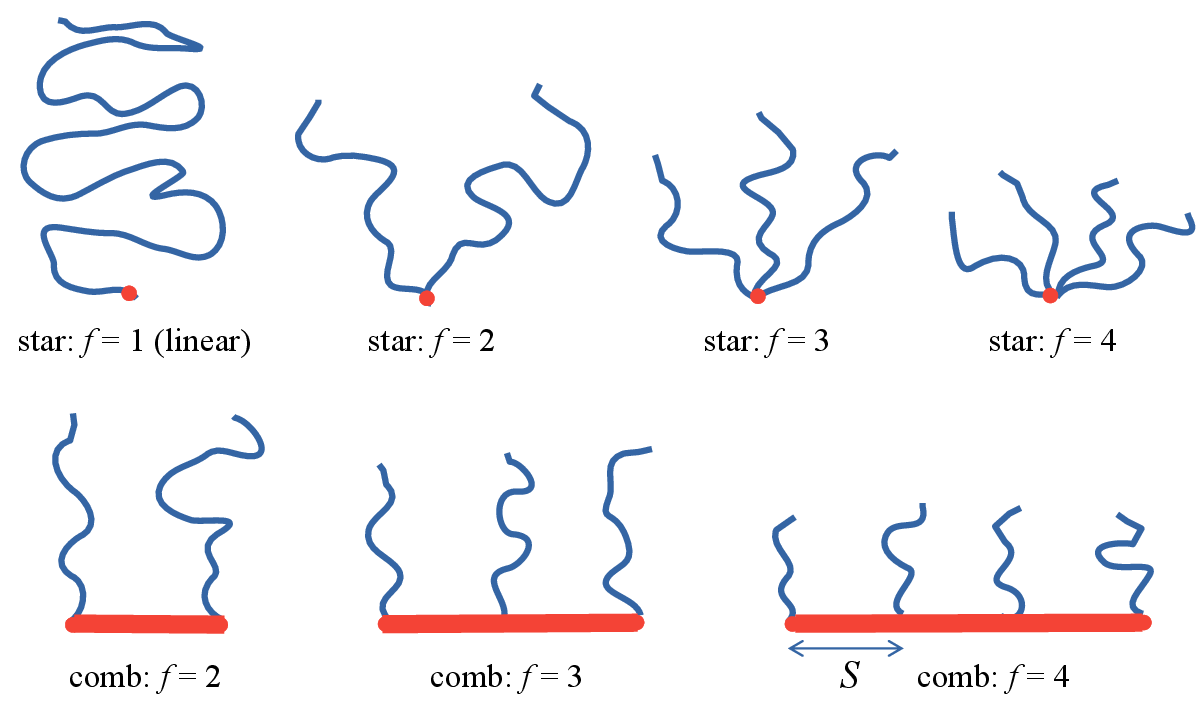}
\caption{\label{polymers_schema} A few examples of polymeric architectures of the same molecular weight, given by the number of adsorbing monomers, $N_\mathrm{total}$, as considered throughout this study. Star- and comb-like architectures with $f=2-4$ ligands are shown, separation between ligands along a backbone of the latter is denoted by $S$.}
\end{figure}
Many branched polymeric architectures can be classified as the core-shell-like structures comprising certain type of a central core which is decorated by a number of ligands \cite{GhoshChaudhuri2011, Kakkar2017, Haino2024}, often displaying super-adsorbent features \cite{Pastrafidou2025}. 
Such structures can be classified by the dimensionality of a core, $d_c$, adsorbing ligands length, $N$, and their number, $f$, where the latter defines the density of a shell. Two simplest cases are: zero-dimensional core, $d_c=0$, mimicking a star-like polymer \cite{Schaefgen1948, Ren2016, Liu2017, Jo2022, Shin2024, Yi2024}; and the one-dimensional core, $d_c=1$, yielding a comb-like structure with a rigid linear backbone \cite{Roovers1975, Lipson1987, Radke2005}, as illustrated in Fig.~\ref{gener_stars}. These two polymeric architectures are considered in this study. Experimental examples of a rigid backbone of a comb-like polymer include linear isomers of higher phenyls, polysaccharide, polystyrene, etc. The cases with higher dimensionality of cores include: $d_c=2$ with the examples being the core of discotic liquid crystals \cite{Bisoyi2010, Wohrle2015, Termine2021} and of polycyclic heteroaromatic compounds \cite{Stepien2016}; $d_c=3$, with a cubic \cite{Lacroix2011, Singh2022, Mitzi2019} or near-spherical \cite{Umar2021, Tanzi2022, Alberti2022} core shapes of metal nanoparticles or organic structures. These could be potentially considered in the future.

The focus of the current study is to quantify the effect of polymer branching, characteristic to the star-like and comb-like polymeric architectures, on their ability to adsorb obstacles, using computer simulations. Given the separation between adjacent chains in the comb-like structure is constant and equal to $S$, the limit of $S=0$ degenerates into the case an $f$-branch star. We assume all $f$ linear adsorbing branches to be of the same length, $N$, and a core to be composed of non-adsorbing monomers. Then the molecular weight of the adsorbing part of any branched polymer is defined via $N_{{\rm total}}=fN$ allowing comparison between the cases of $d_c=0$ and $1$ on equal terms. Several examples of such polymeric architectures, with various branching parameters, $f$ and $N$, but the same $N_{{\rm total}}$, are shown in Fig.~\ref{polymers_schema}.

Besides global properties characterizing adsorption, we will also analyze the statistics of the effect of bridging between two different chains via shared obstacle(s). This will clarify whether this effect plays any profound role in increasing of adsorption efficiency of branched adsorbent as discussed in Refs.~\cite{Rosthauser1981, Samuel2017, Zhang2019, Johann2019, Wurm2020, Tyagi2020, Mouhamad2023}.   

To this end we will employ macroscopic description of branched polymer adsorbents based on a discrete model of growing polymers on regular lattices \cite{deGennes1979, desCloizeaux1982, Freed1981, Hooper2020}, which proved its high efficiency in retrieving universal conformational statistics of polymers in good solvent regime, and has been successfully applied in our previous studies \cite{Blavatska2024a, Blavatska2024b} for the case of linear polymer adsorbents. Within this approach, the obstacles are defined as a set of special lattice sites characterized by an effective short-range attraction to monomers of a polymer structure. We will consider two possible scenarios: i) the regime of immobilized obstacles, where various polymer conformations are realized in an ensemble of various frozen random distributions of obstacles on a lattice, and ii) the regime of diffusive obstacles, where a polymer structure serves as a flocculant/chelant for obstacles performing Brownian motion, thus enabling the analysis of the obstacle-polymer adsorption dynamics. 

The layout of the rest of the paper is as follows. In Section \ref{II} we introduce the lattice polymer growth algorithm and apply it to build-up the ensembles of the core-shell polymer structures. Our main results on quantitative description of adsorption on the star-like and comb-like structures are presented in Section \ref{III}, followed by Conclusions.

\section{The model and the method} \label{II}

We introduce the $3D$ simple cubic lattice of $L^3$ sites, where we construct required core-shell-like polymer structures, via polymer growth algorithm. For the case of $d_c=1$ with an uniform separation, $S$, the linear core has a form of a rigid rod with the length of $S(f-1)$, collinear with the $OX$ axis. Its starting point is chosen at the middle of a lattice, given by the coordinates $\{ x^c_1, y^c_1,z^c_1\}=\{L/2,L/2,L/2\}$. Hence, the coordinates of the core sites are: $\{ x^c_i, y^c_i,z^c_i\}=\{L/2+(i-1),L/2,L/2\}$, for all $i=2,\ldots,S(f-1)+1$. For the case of zero separation, $S=0$, the core degenerates into a single point with coordinates $\{ x^c_1, y^c_1,z^c_1\}$ mimicking the case of $d_c=0$ (a star-like polymer). Note that within our approach the core does not have adsorption properties and serves purely as a scaffold for grafting branches.

\begin{figure}[!h]
\vspace{-4em}
\includegraphics[width=80mm]{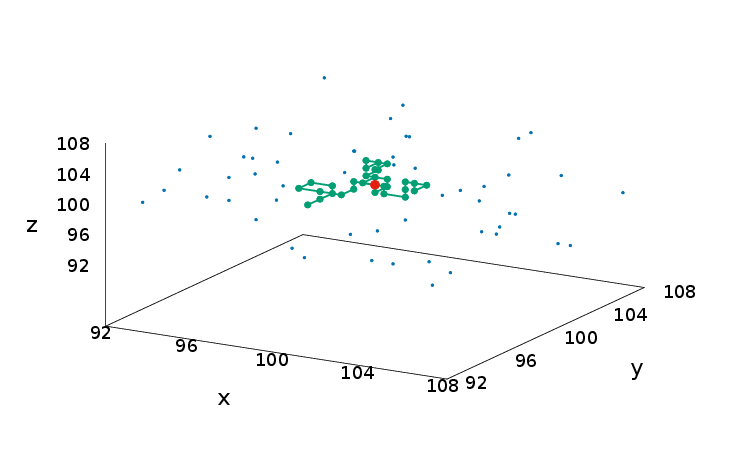}
\vspace{-4em}\\
\includegraphics[width=80mm]{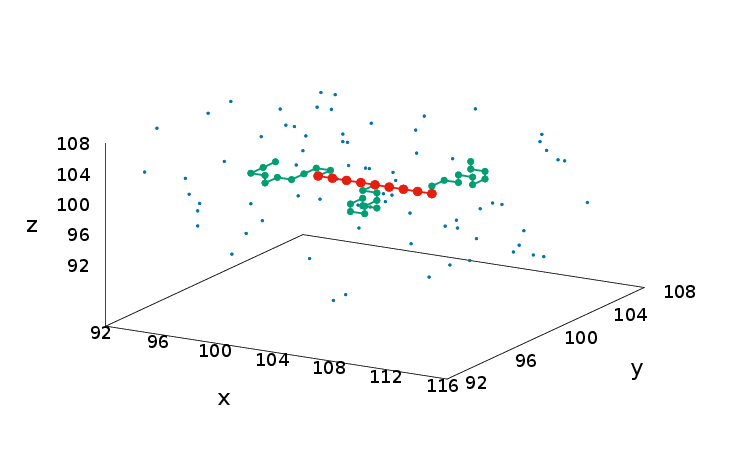}
\vspace{-2em}
\caption{\label{starscheme}Snapshots of two core-shell-like structures: $d_c=0$, $f=3$ (top) and $d_c=1,f=3,S=4$ (bottom) immersed into the lattice with obstacles (as shown by small blue dots). The lattice sites belonging to the cores are shown with red symbols, while those occupied by monomers are displayed in green.}
\end{figure}

The required number, $f$, of linear branches, see Fig. \ref{polymers_schema}, are to be built next. We apply the pruned-enriched Rosenbluth Method (PERM) \cite{Grassberger97}, based on the  Rosenbluth-Rosenbluth (RR) algorithm of growing chain \cite{Rosenbluth55} along with enrichment strategies \cite{Wall59}. Each of the $f$ linear polymer chains (branches) is grown step-by-step, i.e., each new, $n$th, monomer is introduced at a randomly chosen unoccupied nearest neighbor site to the previous, $n-1$, monomer satisfying self-avoidance of a chain, until the required total number of monomers, $N$, is reached. The coordinates of each $n$th monomer ($n=1,\ldots,N$) of $k$th branch ($k=1,\ldots,f$) are designated as $\vec{R}_n^{k}=(x_{n}^{k}, y_{n}^{k}, z_{n}^{k})$, where $|\vec{R}_n^{k}-\vec{R}_{n-1}^{k}|=1$. The first monomer $n=1$ of each linear branch is introduced at a randomly chosen unoccupied nearest neighbor site of respective grafting point on a core.  Note that the sites belonging to core are avoided by the monomers of growing chains. The examples of such lattice core-shell structures are given by the snapshots shown in Fig.~\ref{starscheme}.   

A weight $W_n$ is assigned to each configuration at the $n$th step, given by the Rosenbluth weights \cite{Rosenbluth55}
\begin{equation}
W_n^{{\rm Rosenbluth}}= \prod_{l{=}2}^n m_l, \label{weight}
\end{equation}
where  $m_l$ is the number of all unoccupied lattice sites available for adding the $l$th monomer.
Population control in PERM suggests pruning configurations with too low weights, and enriching the sample with the copies of the high-weight configurations \cite{Grassberger97} by a proper choice for the threshold weights, $W_n^{<}$ and $W_n^{>}$. If the weight $W_n$ of a running conformation of a chain is less than $W_n^{<}$, it is discarded with the probability $1/2$, otherwise it is kept and its weight is doubled. If $W_n$ exceeds  $W_n^{>}$, the configuration is doubled and the weight of each copy is taken as half the original weight. 

The thermodynamic averaging for any observable of interest $O$ is thus given by
\begin{equation}
 \overline {O } =\frac{{\sum_{{\rm conf}} W_N^{{\rm conf}} O }} {Z_N} \label{confaver}
 \end{equation}
 with  $W_N^{{\rm conf}}$ being the weight of an $N$-monomer chain in a given configuration given by (\ref{weight}) and  ${Z_N} =\sum_{{\rm conf}} W_N^{{\rm conf}}$ is the partition sum.

After the growth of a polymer structure is completed, the fraction $p$ of adsorbant obstacles are introduced at randomly chosen unoccupied sites of the lattice. As discussed in the Sec.~\ref{I}, these are interpreted as microflocs, heavy metal ions, etc., assuming that the lattice constant is comparable with the characteristic dimension of a particular obstacle. We label their coordinates, $\{X(i),Y(i),Z(i)\}$, by the function $s(X(i), Y(i), Z(i))$ such that
$$
s(X(i), Y(i), Z(i)){=}\begin{cases}
1, \, \,   \text{if $i$th site contains an obstacle,}\\
0, \, \,   \text{otherwise. }
\end{cases} 
$$
If one or more of the nearest-neighbor sites of $i$th obstacle is occupied by a polymer, this obstacle is assumed to be adsorbed on it. This models a short-range attraction between the polymer and obstacles, and each such monomer-obstacle contact will be interpreted hereafter as a single adsorption bond for $i$th obstacle. 

Note that double averaging is performed for each characteristic $O$: the disorder averaging
\begin{equation}
  \langle {\overline {O}} \rangle= \frac{1}{M}\sum_{i=1}^M {\overline {O}}^i \label{aver}
 \end{equation}
is performed over an ensemble of $M=1000$ different realizations (replicas) of random spatial arrangement of obstacles on top of the thermodynamic averaging, ${\overline{O}}^i$, in $i$th replica of a lattice, according to (\ref{confaver}). 

The main observables of interest are introduced next. 
\begin{enumerate}
    \item 
The total number, $N_s$, of adsorption bonds for given realization of the polymer structure and given distribution of obstacles is calculated as:
\begin{eqnarray}
%&&N_s=\sum_{n=1}^N\sum_{k=1}^f\sum_{g=1}^G %\Big(s(x_n^{g,\,k}\pm1,y_n^{g,\,k},z_n^{g,\,k})+\nonumber\\
%&&+s(x_n^{g,\,k},y_n^{g,\,k}\pm1,_n^{g,\,k})+s(x_n^{g,\,k},y_n^{g,\,k},z%_n^{%g,\,k}\pm1)\Big). \label{Nsdef}
&&N_s=\sum_{n=1}^N\sum_{k=1}^f \Big(s(x_n^{k}\pm1,y_n^{k},z_n^{k})+\nonumber\\
&&+s(x_n^{k},y_n^{k}\pm1,_n^{k})+s(x_n^{k},y_n^{k},z_n^{k}\pm1)\Big). \label{Nsdef}
\end{eqnarray}
The normalized average value $  {\overline  {\langle n_s \rangle}}=  {\overline  {\langle N_s  \rangle }}/N_{{\rm total}}$  {provides adsorption capacity of a single monomer in terms of adsorption bonds.}

\item
Another parameter is the total number, $N_a$, of obstacles adsorbed onto a given polymer structure.
To this end, we again prescribe the label function  $b(X(i),Y(i),Z(i))$ to each $i$th site of the lattice containing an obstacle, with initial values $b(X(i),Y(i),Z(i))=0$.  
We are then checking the conditions, for $n=1,\ldots, N$ and for $k=1,\ldots, f$:
 \begin{itemize}
     \item 
 if $ s(x_n^{k}\pm1,y_n^{k},z_n^{k})=1$, then 
$ b(x_n^{k}\pm1,y_n^{k},z_n^{k})=b(x_n^{k}\pm1,y_n^{k},z_n^{k})+1$;
     \item 
 if $ s(x_n^{k},y_n^{k}\pm1,z_n^{k})=1$, then 
$ b(x_n^{k},y_n^{k}\pm1,z_n^{k})=b(x_n^{k},y_n^{k}\pm1,z_n^{k})+1$;
 \item 
 if $ s(x_n^{k},y_n^{k},z_n^{k}\pm1)=1$, then 
$ b(x_n^{k},y_n^{k},z_n^{k}\pm1)=b(x_n^{k},y_n^{k},z_n^{k}\pm1)+1$;
 \end{itemize}
so that at the end the values of $b(X(i),Y(i),Z(i))$ contain the total number of bonds of an obstacle positioned at site $i$ with monomers of polymer structure. Thus, the number of different obstacles  $N_a$ encountered by monomers of polymer structure is given by
 \begin{eqnarray}
&&N_a=\sum_{n=1}^N\sum_{k=1}^f \Big(\delta_{b(x_n^{k}\pm1,y_n^{k},z_n^{k}),1}+\nonumber\\
&&+\delta_{b(x_n^{k},y_n^{k}\pm1,z_n^{k}),1}+\delta_{b(x_n^{k},y_n^{k},z_n^{k}\pm1),1}\Big), \label{Nadef}
\end{eqnarray}
where $\delta$ is the Kronecker delta, i.e. the summation is performed over those obstacles that have been encountered only once. Again, the normalized average value, $ {\overline  {\langle n_a  \rangle}}=  {\overline  {\langle  N_a  \rangle }}/N_{{\rm total}}$,  {provides adsorption capacity of a single monomer in terms of the average number of obstacles adsorbed by it.}

\item The average number of bonds, $n_{{\rm bond}}$, established by a single obstacle with monomers can then be easily evaluated as \begin{equation}
{\overline  {\langle n_{{\rm bond}} \rangle}}={\overline  {\langle N_s \rangle}} / {\overline  {\langle N_a \rangle }}. \label{rel}
\end{equation}
This quantity refers to as an average coordination number of adsorbed particles, and can be treated as a measure of  adsorption strength of a given polymer structure: the obstacles attached to a polymer via more bonds are less likely to be desorbed by thermal fluctuations or a flow, in case these will be introduced into the model.
\end{enumerate}

Since these three main characteristics of interest are interrelated through Eq. (\ref{rel}), we will mainly focus in what follows on two of them, 
${\overline  {\langle n_{{a}} \rangle}}$ and
${\overline  {\langle n_{{\rm bond}} \rangle}}$.

\section{Results} \label{III}

\subsection{The case of immobilized obstacles}

We consider the case of immobilized obstacles here, but the ensemble averaging is performed over various realization of their spatial arrangements. The latter can be interpreted as snapshots of the impurities distributions at different time instances. Therefore, the averaging still accounts in an implicit way for chaotic movements of impurities. {It is quite obvious that the prerequisite for the polymer to adsorb obstacles is the presence of some number, $N_{\mathrm{free}}(i)$, of unoccupied, ``free", lattice sites within the nearest neighborhood of each $i$th monomer. Statistics of such free sites reflects the ``accessibility" of individual monomers to serve as adsorbing sites and, therefore, is of much interest in the context of our study. For this purpose it is sufficient to consider the polymer structures only, with no impurities. We split such our analysis of polymer conformations and monomers accessibility into two stages.

{At the first stage of our analysis, we focus on how such accessibility depends on the topological distance of an individual monomer from a core. To this end we introduce the sequential numbering of the monomers along each $k$th branch ($k=1,\ldots,f$) as $n_k=1,\ldots,N$.  The number $N_{\mathrm{free}}(n_k)$ of  free neighbor sites around each monomer, $n_k$, is evaluated and the mean  over all $f$ branches: $n_{\mathrm{free}}(n)=\sum_{j=1}^fN_{\mathrm{free}}(n_j)/f$.}

\begin{figure}[!h]
 \begin{center}
\includegraphics[width=8.2cm]{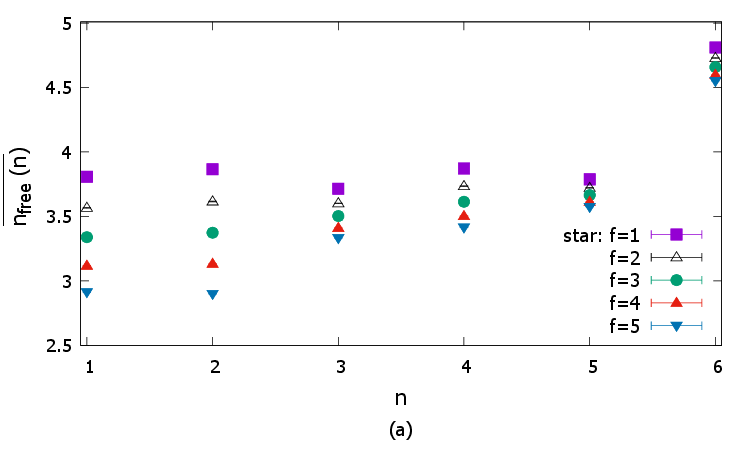}
\includegraphics[width=8.2cm]{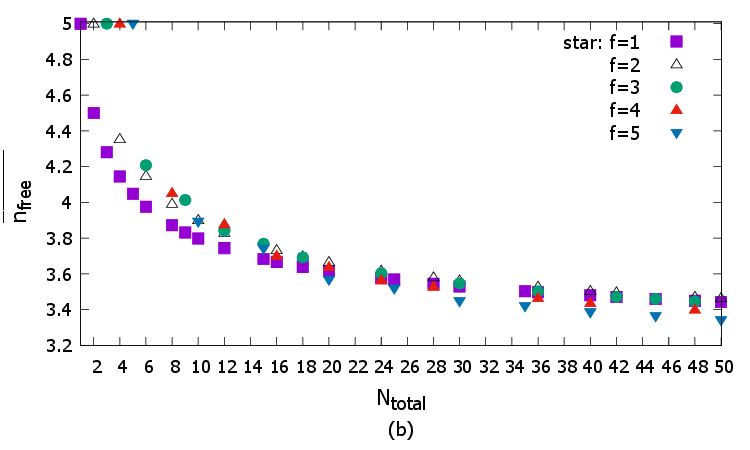}
\end{center}
{\caption{\label{nfree} a) Average number of ``free'' nearest neighbor sites around the monomers with the topological distance $n$ from the branching point of star-like structures (at $N=6$). b) The total accessibility ${\overline {n_{\mathrm{free}}}} $  of a star-like structure as a function of its total molecular weight at different $f$. }}
\end{figure}

{The result of the ensemble averaging for all star-like polymers containing $f=1,\ldots,6$ branches each of the length $N=6$, is shown in Fig.~\ref{nfree}a. For any $f$, the furthest monomer, $n=6$, is the most accessible comparing with the monomers nearer to the core. There are two obvious reasons for this: (i) the end monomer has maximum five free neighboring sites, whereas the internal ones have only four; and (ii) internal monomers are subject to crowdedness effect because of the presence of a common branching site for all branches.} As the consequence of (ii), with increasing $f$ from $1$ to $5$, the innermost monomers, $n=1$ and $2$, become gradually less accessible and their probability to meet the obstacle particles is decreasing. This reflects the idea of Daoud and Cotton \cite{Daoud1982} upon the presence of a dense core in the vicinity of a star-like polymer center.

\begin{figure}[!h]
 \begin{center}
\includegraphics[width=8.2cm]{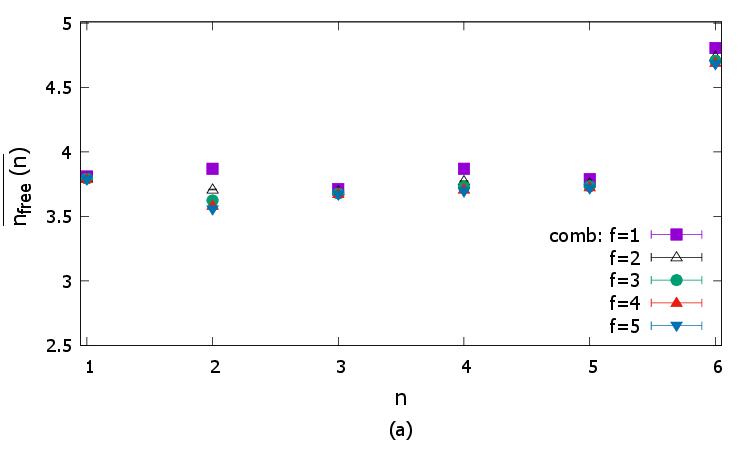}
\includegraphics[width=8.2cm]{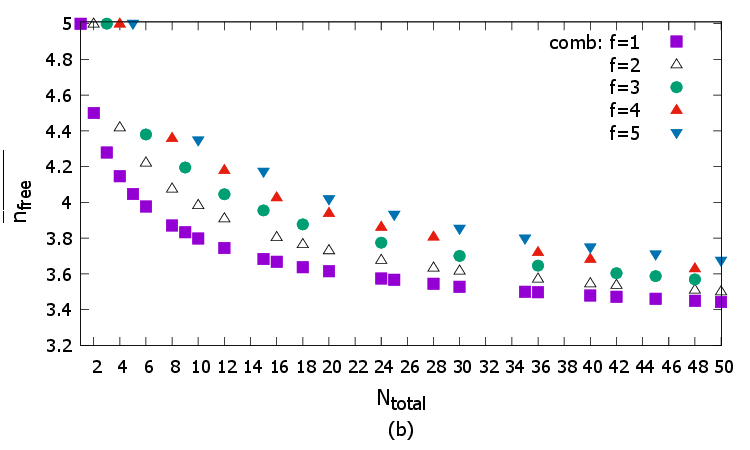}
\end{center}
{\caption{\label{nfreecomb}  a) Average number of ``free'' nearest neighbor sites around monomer with position $n$ along the single branch of comb-like structures with $N=6$, $S=4$. 
b) The total accessibility ${\overline {n_{\mathrm{free}}}} $ of  the comb-like structure with  $S=4$ as function of its total molecular weight at different $f$.}}
\end{figure}

{At the second stage of the analysis, we introduce the number of accessible nearest neighbor sites  per one monomer averaged over all types of monomers of a polymer structure.} It is defined as ${\overline {n_{\mathrm{free}}}}=\sum_{m}{\overline {n_{\mathrm{free}}(m)}}/N$, and characterizes the accessibility of the branched macromolecule in a whole. This quantity is shown in Fig.~\ref{nfree}b at various $f$ and $N_{{\rm total}}$. As it is intuitively clear, the mostly accessible are the so-called dimer, trimer, tetramer and similar structures with respective number of branches, $f=2,3,4$, and their length fixed at $N=1$. In this case $n_{\mathrm{free}}=5$. In the regime of larger $N$, the accessibility of branched structure gradually becomes lower comparing with linear chain of the same molecular weight, and decreases with an increase of $f$. {We note that at higher molecular weight, $N_\mathrm{total}>16$, the data obtained at various number of branches, $f$, practically overlap, indicating no gain in the increase of the molecular weight in this regime.}

\begin{figure}[!h]
 \begin{center}
\includegraphics[width=8.2cm]{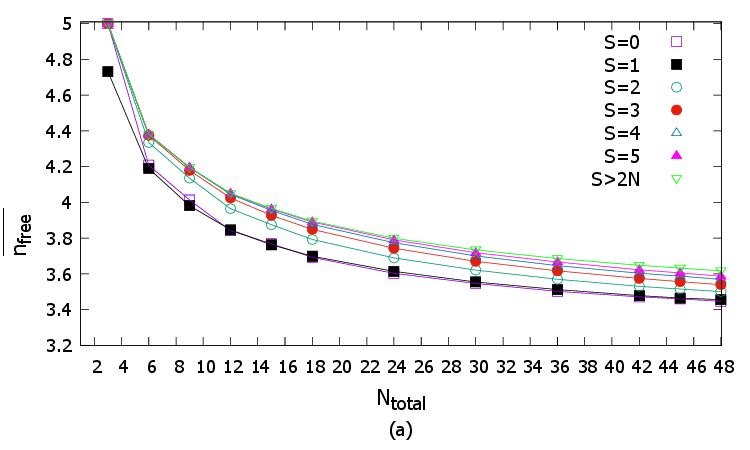}
\includegraphics[width=8.2cm]{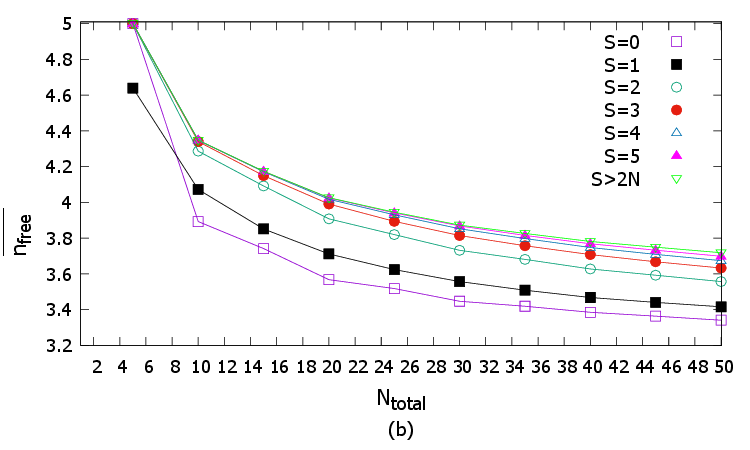}
\end{center}
{\caption{\label{nfreenum35}  
The average accessibility of comb-like polymer structures with $f=3$ (a) and $f=5$ (b) at various $S$ as functions of total number of monomers.}}
\end{figure}

The same set of properties is also analyzed for the comb-like structures. We evaluate the values for  ${\overline{n_{{\rm free}}(n)}}$, averaged over an ensemble of various polymer configurations with $f=1,\ldots,5$ for a certain characteristic case with $S=4$ and $N=6$, see Fig.~\ref{nfreecomb}a. The case $f=1$ retrieves a single linear chain. We notice that the monomers adjacent to the core, $n=1$, are equally accessible irrespective of $f$, in contrast to the case of a star-like structure (cf. Fig. \ref{nfree}a). This is the consequence of the nonzero separation $S=4$ between the grafting on a core, in which case the adjacent branches do not mutually interfere the accessibility of their monomers. The monomers further away from a core, $n>1$, are still a bit less accessible comparing with the case of completely independent chains ($S\to\infty$), but the crowdedness effects are much less pronounced already at the separation of $S=4$  (cf. the case of $S=0$ in Fig.~\ref{nfree}a).

The total averaged accessibility, ${\overline {n_{\mathrm{free}}}}$, of such comb-like structures at fixed $S=4$ is presented in Fig.~\ref{nfreecomb}b. At any given $f>1$ and $N$, the accessibility of such structure is higher comparing with those of linear chain of the same molecular weight and this effect is well pronounced, in contrast to the case of star-like structures as shown in Fig.~\ref{nfree}b.

\begin{figure}[htb]
 \begin{center}
\includegraphics[width=8.2cm]{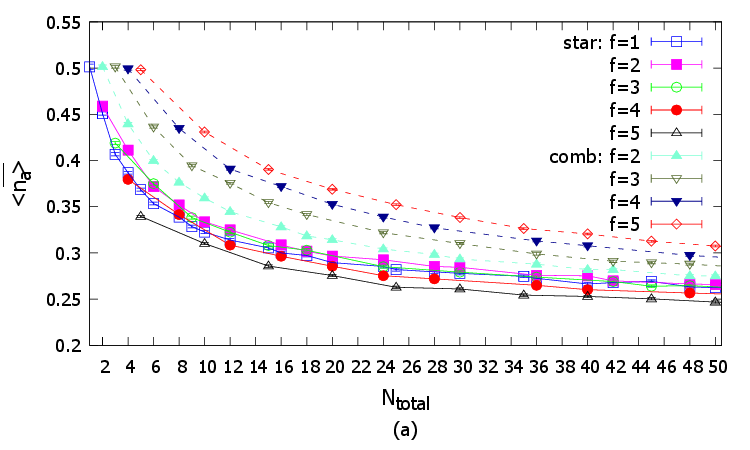}
\includegraphics[width=8.2cm]{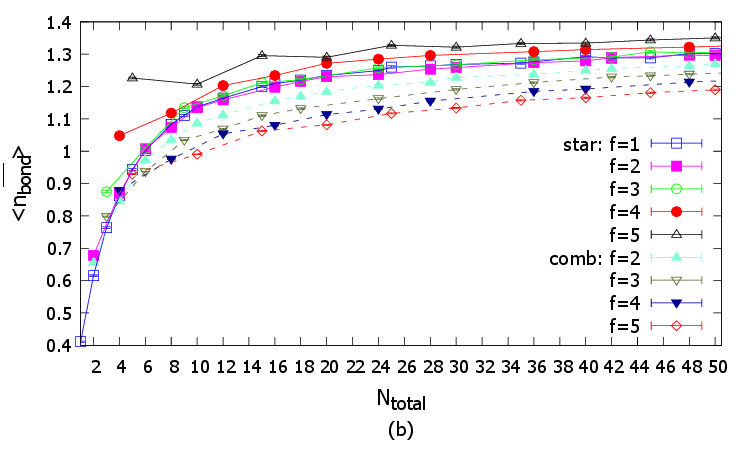}
\end{center}
{\caption{\label{nan_all} a) Adsorption capacity given by $\langle {\overline{n_{a}}} \rangle$, and b) an average coordination number given by  $\langle {\overline{n_{\rm bond}}} \rangle $ as the functions of $N_{{\rm total}}$ at concentration $p=0.1$. The cases of star-like structures  (solid lines) and comb-like structures with fixed $S=4$ (dashed lines) are shown.
}}
\end{figure}

It is obvious that the excluded volume interference between the chains of a comb-like structures depends on the separation between their grafting points, $S$. Two limits exist here: $S=0$, a star-like polymer with maximal possible interference between chains; and $S\to\infty$, the limit of independent chains. For the intermediate values of $S$ the average monomer accessibility, ${\overline {n_{\mathrm{free}}}}$, is a monotonic function of $S$ as shown in Fig.~\ref{nfreenum35}. Here, the average accessibilities are compared side-by-side between the star-like and comb-like topologies, at various $S$ and at fixed number of branches: $f=3$ and $f=5$. It is worth noting that while at any value of $S$ the mostly accessible are trimer and pentamer structures with $N_{{\rm total}}=3$ and $5$, respectively, in which case ${\overline {n_{\mathrm{free}}}}=5$; this does not hold for a special case of $S=1$, when ${\overline {n_{\mathrm{free}}}}<5$. As expected, in the regime of larger $N_{{\rm total}}$ increasing of $S$ results in increase of accessibility values at each fixed $N_{{\rm total}}$ because of the reduction of the crowdedness effect and moving towards the regime of independent branches. This regime is practically achieved already at $S>2N$, and in this case ${\overline {n_{\mathrm{free}}}}$ values reach their maximum for each $N_{{\rm total}}$.     

Monomers accessibility, ${\overline {n_{\mathrm{free}}}}$, evaluated from purely steric effects of a polymer structure as discussed above, is related to its ability to adsorb obstacles. Indeed, this characteristic defines the average numbers of bonds established by a single monomer with adsorbed particles at given concentration $p$ of the latter: ${\overline  {\langle n_{{s}} \rangle}} =p\, {\overline {n_{\mathrm{free}}}}$. In the saturation regime (discussed in the following section), when particles are allowed to diffuse and finally be adsorbed onto a polymer in all possible ways, the maximum value of ${\overline {n_{s}}}$ is reached: ${\overline{\langle n_{{s}} \rangle}} ={\overline {n_{\mathrm{free}}}}$.  

\begin{figure}[h!]
 \begin{center}
\includegraphics[width=8.2cm]{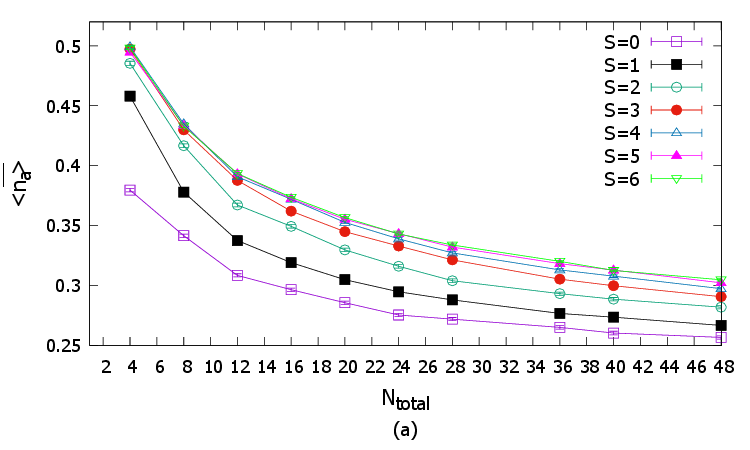}
\includegraphics[width=8.2cm]{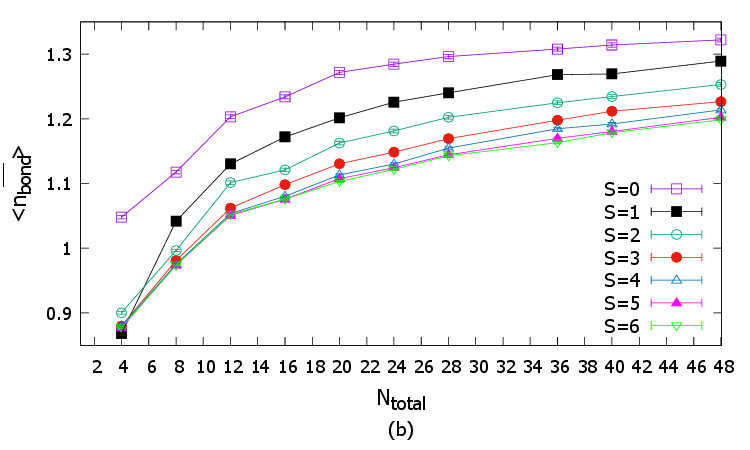}
\end{center}
{\caption{\label{nanum4} a) Adsorption capacity given by $\langle {\overline{n_{a}}} \rangle $,  and b) an average coordination number given by  $\langle {\overline{n_{\rm bond}}} \rangle $  as functions of $N_{{\rm total}}$ at fixed $f=4$ and various $S$.
}}
\end{figure}

We switch to the analysis of adsorption characteristics of core-shell-like structures in a lattice explicitly, by introducing immobilized obstacles with the concentration $p$. In what follows, we consider the case with fixed concentration $p=0.1$, and this choice is based on the following considerations. On one hand, it is small enough to correlate with real concentrations of impurities in wastewater treatment, especially for the case of heavy metal ions; on the other  hand, it is sufficient to obtain statistics reflecting non-trivial effects for the characteristics considered in this study. Note also, that in our previous study \cite{Blavatska2024a}, where adsorption properties of linear polymer structures was considered, we covered a range of concentrations, $0.01 \leq p \leq 0.1$, and found rather trivial linear dependence of adsorption characteristcs on $p$. For the sake of convenience, in what follows, we will focus on presenting the results obtained at various $f$ and $S$ at fixed total molecular weight $N_{\rm total}$ (let us remind that one can easily extract the value of the length of individual branches as $N=N_{\rm total}/f$). In such a format, one can easily discriminate the impact of branching, in terms of $f$ and $S$, on the adsorption characteristics of different polymer topologies with the same molecular weight.

The results obtained for adsorption capacities, ${\overline {\langle n_{{a}} \rangle}}$, for the set of star- and comb-like structures are presented in Figs.~\ref{nan_all}a and \ref{nanum4}a for a range of parameters $f$ and $S$. In all situations considered, the structures with shorter branch lengths, $N_{\rm total}$, are found to be more efficient in adsorbing obstacles than the ones with longer branches. This is directly related to the same tendency of average accessibility as explained before (cf. Figs.~\ref{nfree}b and \ref{nfreecomb}b). For the case of star-like structures, increasing the number of branches $f$ at each fixed $N_{{\rm total}}$ leads to decreasing of adsorption capacity as comparing with the single linear chain of the same total molecular weight. This is explained by reduction of their effective sizes (e.g. well-known compactification effect manifested by the decrease of the average gyration radius of a star comparing to its linear counterpart of the same molecular weight \cite{Zimm1959}, due to mutual crowdedness of branches). This shrinks available unoccupied neighborhood of its monomers for adsorbing new obstacles. However, the comb-like structures with $S>0$ become more efficient in adsorbing new obstacles with increasing the number of branches $f$: effect of crowdedness of individual branches is much less pronounced at large $S$, which allows the branches to extend in space and be more able to adsorb the obstacles. As already discussed above, the limit of large $S$ represents a set of independent linear branches (Fig.~\ref{nanum4}a). 

The average number of bonds (contacts) ${\overline  {\langle n_{{\rm bond}} \rangle}}$ between monomers and adsorbed obstacles, evaluated per single obstacle (averaged coordination number of obstacles) is addressed next. It quantifies stability of a macrofloc (formed by a polymer and adsorbed microflocs), or of the chelation complex with the metal ion. This characteristic is plotted in Figs.~\ref{nan_all}b and \ref{nanum4}b at a range of the values for $f$ and $S$. For the case of star-like structures, increasing the branching parameter $f$ at each fixed $N_{{\rm total}}$ leads to an increase of average coordination number. Indeed, compactification of effective size of star polymers  due to mutual crowdedness of branches leads to increasing probability to form multiple bonds with a single obstacle. Opposite picture is observed for comb-like structures with a rod-like core:  with increasing $S$, the effective size of such structures becomes more extended in space and their abilities to establish multiple contacts with already adsorbed obstacles becomes smaller (Fig. \ref{nanum4}b).

An important aspect related with mutual crowdedness of the branches of star-like structures ($S=0$) and its gradual extinction with increasing of  branch separation $S$ concerns the effect of bridging between different branches of core-shell polymer structures via shared absorbed particle(s). Indeed, the single particle can establish multiple bonds not only with the monomers belonging to the same branch of polymer structure, but also to monomers of different branches simultaneously. In such a way, the adsorbed particles serve as ``bridges'' connecting two or more different branches \cite{Tyagi2020}.

The average probability for a single particle to establish simultaneous bonds with $g$ branches ($g=1,\ldots,f$), depending on separation between branches $S$, is shown in Fig. \ref{imovbond}. The case of $g=1$  corresponds to the probability for a particle to have bonds with monomers of a single branch only. We notice that for the case of star-like structures ($S=0$), where the crowdedness between the branches is mostly pronounced,  the adsorbed particles establish bridges between two different branches most often, and the probability of bridging between three branches simultaneously is also non-zero in this case, albeit it is rather small. Situation changes with the increase of $S$: the probability for a particle to establish bond with monomers of the same branch starts to increase, due to extinction of mutual crowdedness between branches. This effect is responsible for an overall decrease of average coordination numbers  ${\overline  {\langle n_{{\rm bond}} \rangle}}$  of comb-like structures with increasing $S$, as discussed earlier (cf. Fig. \ref{nanum4}b) and, thus, provides its contribution to the observed impact of branching on the adsorption properties.

\begin{figure}[!h]
\includegraphics[width=8.2cm]{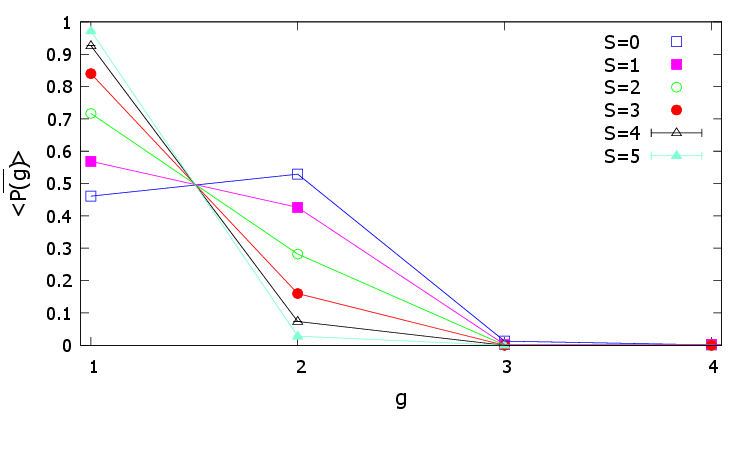}
\caption{\label{imovbond} Probability for a single obstacle to establish bonds with $g$ branches of complex polymer structure at fixed $f=5$, $N_{{\rm total}}=40$ and various $S$. }
\end{figure}

Possible correlation between a typical dimension of a core-shell polymer structure, as provided by its gyration radius, $R_g$, and its adsorption characteristics, $n_a$  and $n_{{\rm bond}}$, is analyzed for three representative cases: single linear chain ($f=1$), star-like structure with $f=4$ and comb-like structure with $f=4$, $S=4$ with the same total molecular weight $N_{{\rm total}}=40$, as shown in Fig. \ref{corr}. Indeed, we observe that the typical values for $R_g$ of a linear chain are larger than these for a star-like structure, but smaller than for a comb-like structure. The largest dimensions of the comb-like structures correlates well with their highest adsorption capacity, $n_a$; as well as with their lowest $n_{{\rm bond}}$, as its extended dimensions imply a low internal density with reduced probability to form the multiple polymer-obstacle bonds.

\begin{figure}[h!]
 \begin{center}
\includegraphics[width=4.23cm]{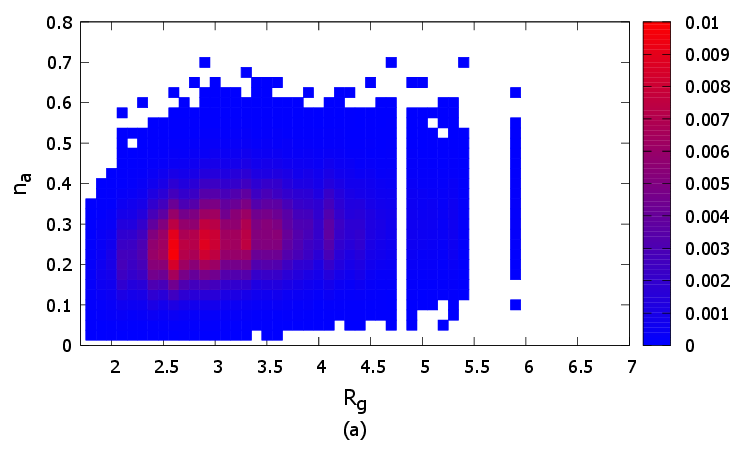}
\includegraphics[width=4.23cm]{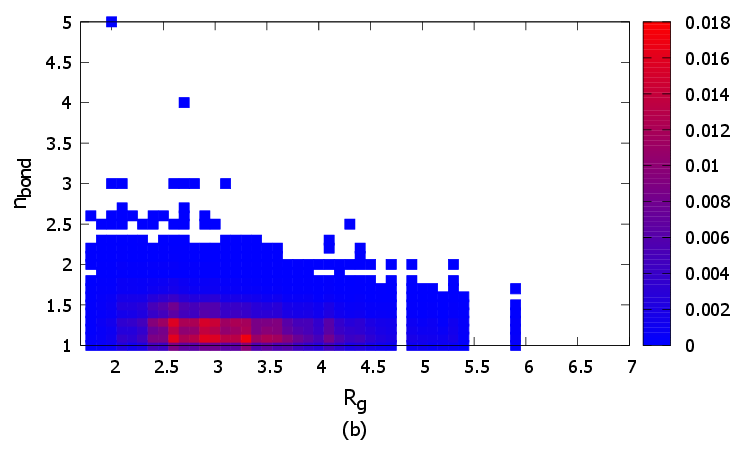}\\
\includegraphics[width=4.23cm]{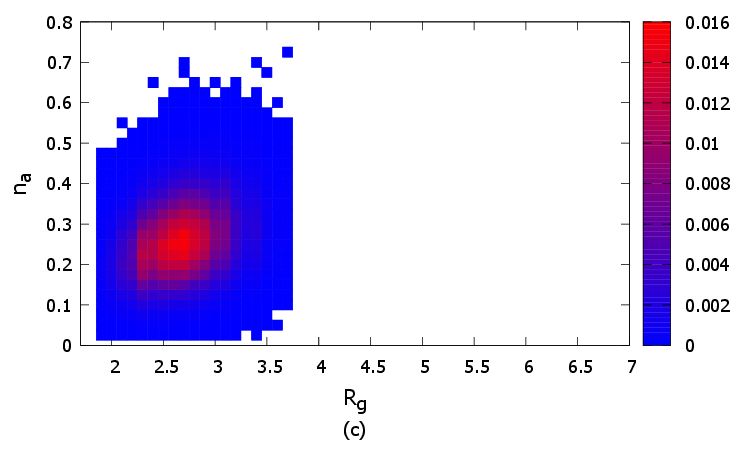}
\includegraphics[width=4.23cm]{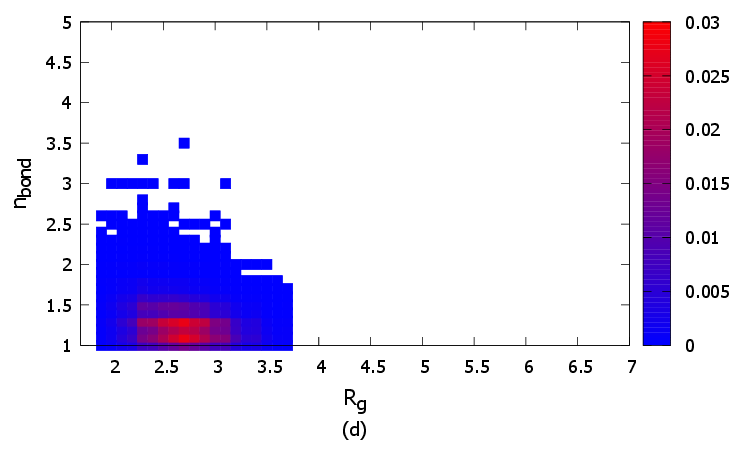}\\
\includegraphics[width=4.23cm]{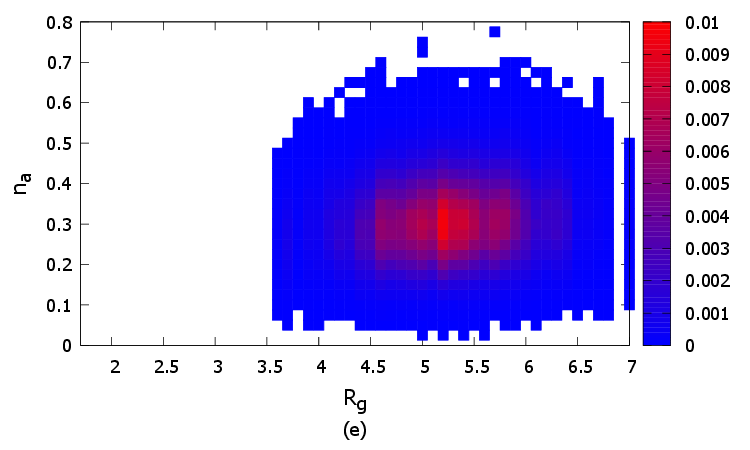}
\includegraphics[width=4.23cm]{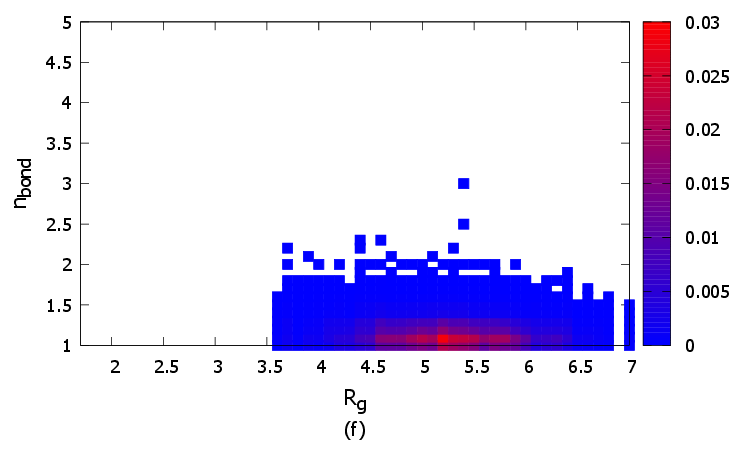}\vspace{-2em}
\end{center}
{\caption{\label{corr} Correlation between the gyration radius $R_g$ of polymer structure and its adsorption characteristics, $n_a$  and $n_{{\rm bond}}$ for a single linear chain (a,b), star-like structure with $f=4$ (c,d), and comb-like structure with $f=4$ (e,f) with the same total molecular weight of $N_{{\rm total}}=40$.   
}}
\end{figure}

We can draw the following conclusions that are based on these results. None of the three polymer structures considered here, a linear chain, star-like, and comb-like topologies, is universally the best for the adsorption of obstacles. The comb-like polymer is the most efficient in adsorbing capacity of obstacles, whereas the star-like polymer is characterized by the strongest retaining abilities via formation of multiple polymer-obstacle bonds. The linear polymer falls inbetween these two cases.

\subsection{The case of diffusive obstacles}

\begin{figure}[!b]
\vspace{-3em}
\includegraphics[width=7cm]{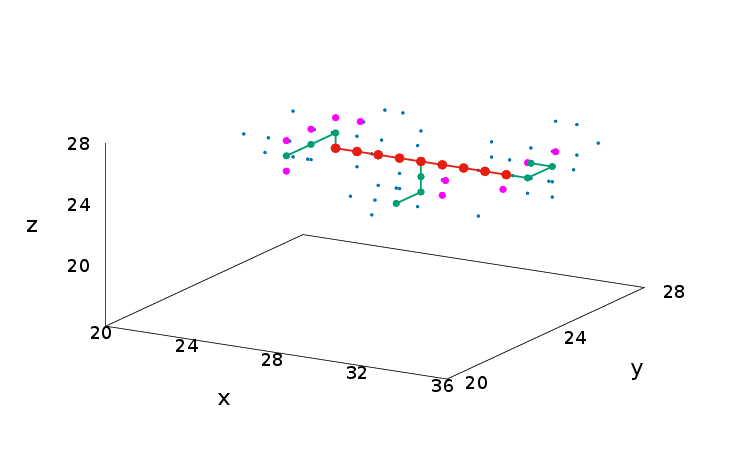}\\
\vspace{-3em}
\includegraphics[width=7cm]{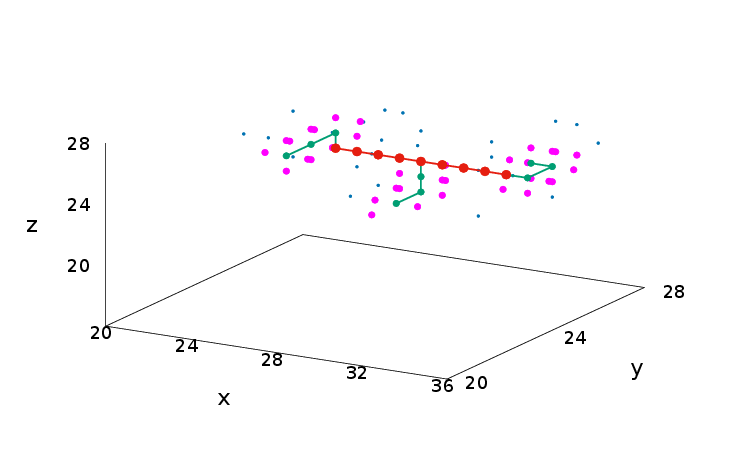}
\vspace{-1em}
\caption{\label{schema_time} Snapshots of comb-like structure with $f=3, S=4$ on the lattice with obstacles (presented with blue symbols) at two time instances: $t=20$ (top) and $t=100$ (bottom). The obstacles adsorbed on a polymer are shown with magenta symbols. }
\end{figure}

{The results presented in the previous section of this study, as well as those discussed in our previous work \cite{Blavatska2024b}, are obtained for the model with quenched obstacles, in which case random diffusion of obstacles is taken into account implicitly, by averaging over their various spatial arrangements. In this section we consider such diffusion explicitly, similarly as in one of our other works \cite{Blavatska2024a}, using suitable Monte Carlo algorithm.}

\begin{figure}[h!]
 \begin{center}
\includegraphics[width=8.2cm]{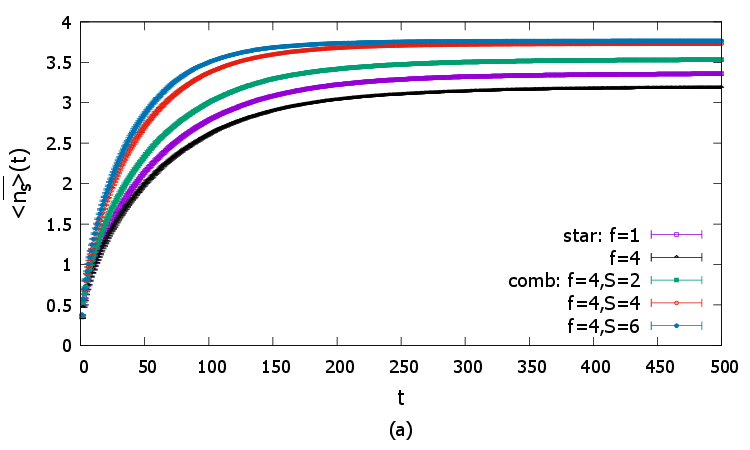}\\
\includegraphics[width=8.2cm]{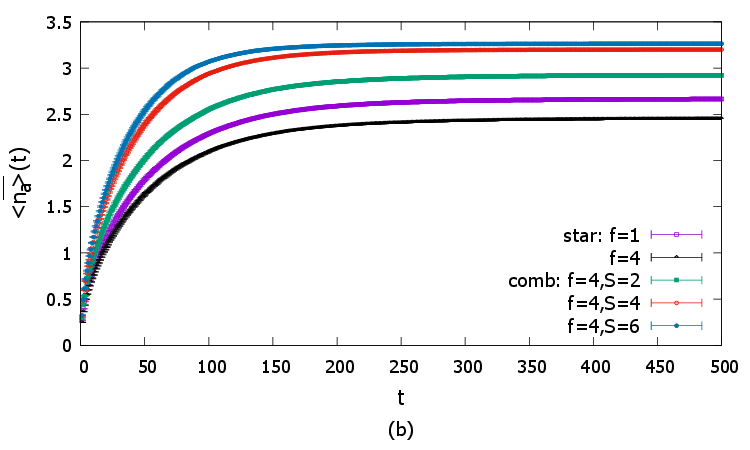}\\
\includegraphics[width=8.2cm]{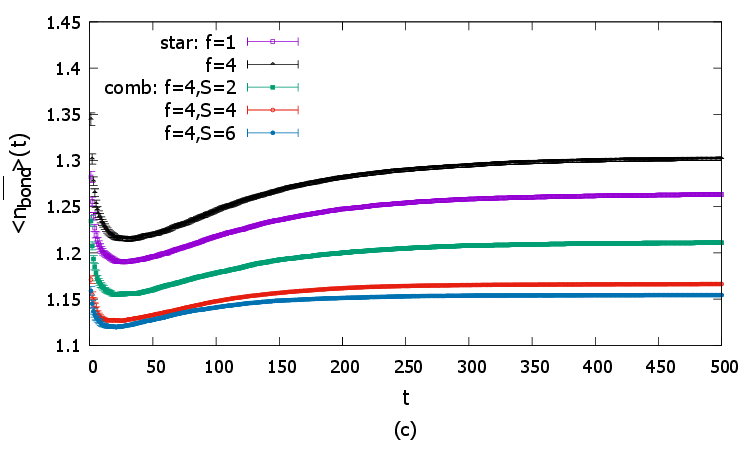}
\end{center}
{\caption{\label{ncrittime} The time evolution for ${\overline  {\langle n_{{s}} \rangle}}$, ${\overline  {\langle n_{{a}} \rangle}}$, and ${\overline  {\langle n_{{\rm bond}} \rangle}}$  (a, b, and c, respectively) for several types of the core-shell-like polymer structures with the same fixed total molecular weight $N_{{\rm total}}=40$.   
}}
\end{figure}

We start from an ensemble of constructed polymer structures, built on a lattice which contains a fraction $p=0.1$ of randomly distributed obstacles, as discussed in the previous section. This initial state, at time $t=0$, is characterized by respective values for adsorption characteristics: $n_s(0)$, $n_a(0)$ and $n_{{\rm bond}}(0)$. The thermal Brownian motion of obstacles is introduced then via a specific Monte Carlo update algorithm which results in the time evolution of these characteristics: $n_s(t)$, $n_a(t)$ and $n_{{\rm bond}}(t)$. The process is schematically illustrated in Fig. \ref{schema_time}). 

We consider the synchronous process, where one time step implies a sweep throughout the whole system. One time evolution step, from $t-1$ to $t$, contains sequential addressing all $i$th obstacles and applying the following rules: 
\begin{itemize}
\item if the obstacle $i$ is already adsorbed onto a polymer structure (one or more of its nearest neighbor sites are occupied by monomers), it simply keeps its current position;
\item if the obstacle $i$ is not adsorbed, it makes a jump towards a randomly chosen unoccupied nearest neighbor site, and if in its new position and obstacle neighbours one or more monomers, it is considered to be adsorbed.
 \end{itemize}
The new spatial positions of obstacles are saved, and after the completion of a sweep their current positions are updated by the new ones. The values $n_s(t)$, $n_a(t)$ and $n_{{\rm bond}}(t)$ are evaluated for the new positions according to their definitions in Section {\ref{II}}, and the time instance is increases by 1.

\begin{figure}[h!]
 \begin{center}
\includegraphics[width=8.2cm]{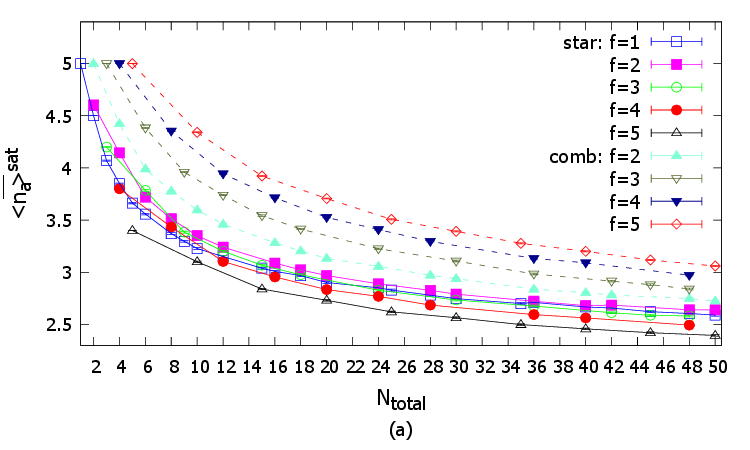}
\includegraphics[width=8.2cm]{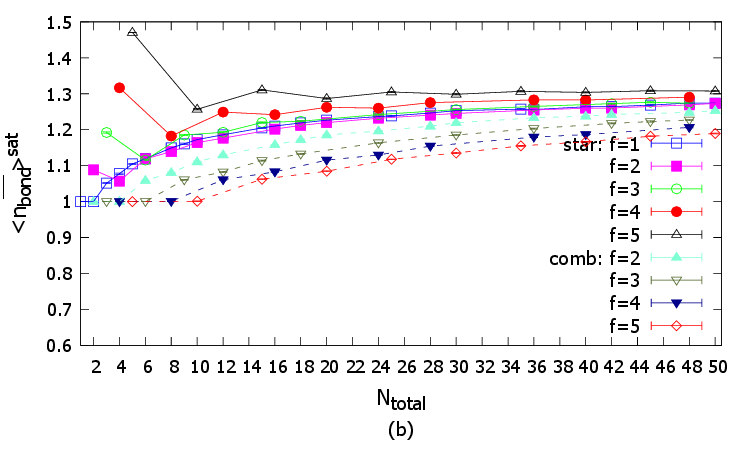}
\end{center}
{\caption{\label{nanfloc} a) Adsorption efficiency given by $\langle {\overline{n_{a}}} \rangle^{{\rm sat}} $  and b) adsorption strength as given by the average number $\langle {\overline{n_{\rm sat}}} \rangle^{{\rm floc}}$  of bonds between a single obstacle particle and monomers in saturation state as functions of $N_{{\rm total}}$. The cases of  star-like structures (solid lines) and comb-like structures with fixed $S=4$ (dashed lines) are considered.
}}
\end{figure}

Typical time dependencies of adsorption characteristics of interest are presented in Fig.~\ref{ncrittime}. The diffusive obstacles are gradually adsorbed on the monomers of polymer structures, and at the very end they form a dense adsorption shell (similar to a solvation shell in a polymer solution) which causes a screening effect for the newly approaching obstacles. Characteristic time required for the number of adsorbed obstacles to reach its saturation value,  $\langle {\overline{n_{a}}} \rangle^{{\rm sat}}$, is $t_{{\rm sat}}$. While we observe the gradual increase of adsorption capacity with $t$ till it reaches the saturation value $\langle {\overline{n_{a}}} \rangle^{{\rm sat}}$, the time dependence of $\langle {\overline{n_{\rm bond}}} \rangle(t)$ demonstrates rather non-trivial behavior with {decreasing regime} at small $t$. We can interpret this as follows. The most accessible perimeter monomers of a polymer structure are the terminal ones, with relatively large number of free neighborhood, $n_{{\rm free}}$. Adsorbing of obstacles on such monomers in the first place, will lead to abundance of single bonds, resulting in decrease of the averaged coordination number $\langle {\overline{n_{\rm bond}}} \rangle(t)$ at early times. With time, the multiply-bonded obstacles, adsorbed to monomers with lower $n_{{\rm free}}$ are starting to make their contributions, leading to an increase of  coordination number towards its saturation value $\langle {\overline{n_{\rm bond}}} \rangle^{{\rm sat}}$.

\begin{figure}[b!]
 \begin{center}
\includegraphics[width=8.2cm]{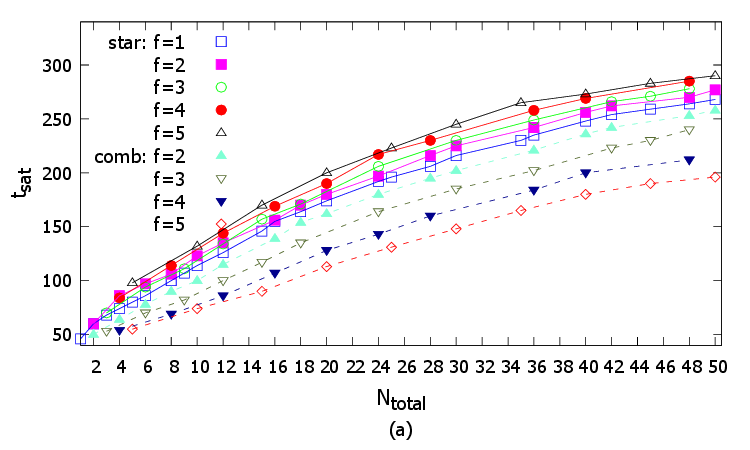}
\end{center}
{\caption{\label{ncrittime2} The  saturation time $t_{{\rm sat}}$   as function of total number of monomers for  the sets of star-like (solid lines) and comb-like (dashed lines) structures with various $f$. 
}}
\end{figure}

The values of $\langle {\overline{n_{a}}} \rangle^{{\rm sat}}$ and $\langle {\overline{n_{\rm bond}}} \rangle^{{\rm sat}}$ in a saturation state are presented in Fig.~\ref{nanfloc}. Let us recall that the case of  $N=1$ (which corresponding to the values for $N_{{\rm total}}$ as high as $5$, depending on the value of $f$), is related to the sets with the highest accessibility of monomers (cf. Figs. \ref{nfree}b, \ref{nfreecomb}b). In the saturation state, the adsorption capacity of comb-like structures in this regime is maximal and is equal to $5$ (\ref{nanfloc}a), whereas the average number of bonds per adsorbed particle is minimal and is equal to 1 $\ref{nanfloc}b$: the obstacles are unable to establish multiple bonding because of the highest accessibility of terminal monomers. For the star-like structures, the situation is opposite:  the values of $\langle {\overline{n_{a}}} \rangle^{{\rm sat}}$ at fixed $N=1$ are decreasing with increasing $f$, while $\langle {\overline{n_{\rm bond}}} \rangle^{{\rm sat}}$   attain the maximum values in this regime (reaching $1.5$ at $f=5$). Apart from this, the general tendency being found for the immobilized obstacles (cf. Figs.~\ref{nan_all}) is preserved: comb-like topologies are more efficient than star-like ones in adsorbing capacity, but the adsorption strength (given via average number of bonds per adsorbed particle) is higher for the more compact star-like structures.

As we already noticed while observing the  time dependences presented in Fig. \ref{ncrittime}, the saturation regime is reached earlier for the cases of comb-like structures, while evolution takes longer time for star-like structures. The rough estimates for $\langle {\overline {t_{{\rm sat}}}} \rangle $  can be obtained e.g. on the basis of data for the time dependence of $\langle {\overline{n_{s}}} \rangle (t)$  under the condition: if $\langle {\overline{n_{s}}} \rangle (t) -\langle {\overline{n_{s}}} \rangle_{{\rm sat}}$ reaches the value $0.05\langle {\overline{n_{s}}}_{{\rm sat}} \rangle$, then $t=t_{{\rm sat}}$. The obtained values for $t_{{\rm sat}}$  as  functions of $N_{{\rm total}}$ are presented in Fig.~\ref{ncrittime2}. The main tendency is again recovering what was observed for averaged accessibilities (cf. Figs.~\ref{nfree}b and \ref{nfreecomb}b): the monomers of comb-like structures are more available for diffusive particles to reach them and to adsorb, compared with those of star-like structures of the same total molecular weight $N_{{\rm total}}$, which results correspondingly in lower values of saturation time.

\section*{Conclusions}

In this study we model two types of processes: coagulation-flocculation of pollutants and chelation of heavy metal ions. Both are widespread in wastewater purification systems and have the same mechanism in common, namely, attachment of some type of a particle (or their agglomerate) to a polymer with a certain structure. The difference in the length scale between these two processes can be brought down by rescaling the problem by a characteristic length of a pollutant and exploiting scaling properties of the polymeric structures \cite{deGennes1979,  desCloizeaux1982}, based on their self-similarity. This brings upon a common paradigm of adsorption of point obstacles onto a polymeric adsorbent, which could be applied to both processes.

Adsorption efficiency of such generalized process can be narrowed down to such factors as: probability of an obstacle to be attached to the adsorbent; adsorbing capacity of adsorbent; and sustainability of an attachment. All these largely depend on the excluded volume effects and  conformational space of an adsorbent, and these, in turn, are defined by its molecular architecture. There is a strong experimental evidence that branched adsorbents are more efficient than their linear counterparts, but it is not quite clear which particular consequences of branching play significant role in this. Here we discuss the effects of branching in the simplest two branched architectures of the core-shell type, namely, a star-like (with a point-like core) and a comb-like (with a linear core) polymers.

To this end we construct a lattice model of these polymers using the pruned-enriched Rosenbluth Method (PERM) \cite{Grassberger97}, based on the  Rosenbluth-Rosenbluth (RR) algorithm of growing chain \cite{Rosenbluth55} along with enrichment strategies \cite{Wall59}. A wide spectrum of star-like and comb-like architectures are covered, each defined via the number of branches (branches), $f$, their length, $N$, and the separation between grafting points, $S$, for the comb-like structures. Following limit cases can be mentioned: a comb-like structure at $S=0$ reduces into a star-like polymer, whereas at $S\to \infty$ this architecture splits into $f$ independent linear chains. 

Two modeling approaches have been considered here: i) the model with immobilized obstacles, where various core-shell polymer conformations are averaged over an ensemble of frozen randomly distributed obstacles, and ii) the model with explicit diffusion of obstacles, with  polymer structures serving as  flocculants/chelants for obstacles performing thermal motion, and dynamics of particles-polymer adsorption process being analyzed.

In the first approach, the quantitative estimates for the set of observables such as the adsorption capacity $  {\overline  {\langle n_s \rangle}}$  per monomer in terms of adsorption bonds, adsorption capacity per monomer in terms of adsorbed obstacles $ {\overline  {\langle n_a  \rangle}}$ and  the average number of bonds $n_{{\rm bond}}$ per particle   (adsorption strength) have been obtained in a wide range of parameters $f$, $N$, and $S$ and  compared between core-shell  topologies with various dimensions of cores at the same total molecular weight   $N_{{\rm total}}$. We found, that the core-shell-like polymer structures with one-dimensional core (comb-like topologies) are more efficient in adsorbing  obstacles (characterised by higher values of $ {\overline  {\langle n_a  \rangle}}$) as comparing with star-like structures of the same total molecular weight, and this effect is more pronounced, the larger is the spatial separation $S$ between grafted branches. On the opposite,  from the point of view of the averaged number of established bonds per monomer   $n_{{\rm bond}}$ 
(adsorption strength), the  star-like structures with zero core are more effective, in particular due to higher probability of bridging between two and more adjacent branches by shared adsorbed particles as compared with the case of more extended comb-like topologies.

In the second approach with explicit diffusion of obstacles, the time dependencies of adsorption characteristics of interest are analyzed. In particular, the quantitative estimates of characteristic time  $t_{{\rm sat}}$ of forming the saturated adsorption shell  of diffusive particles adsorbed onto the monomers of polymer structures is formed, are obtained. The  $t_{{\rm sat}}$ values are  found  to be smaller (and thus the adsorption of diffusive particles is faster) for the case of comb-like polymer structures, since they are more extended and space and thus more available for diffusive particles to reach them and to adsorb, as compared with more compact star-like structures of the same total molecular weight.

Based on these findings one may deduce, that neither the star-like nor comb-like architectures is absolutely the most efficient adsorbent. In particular, the former tops in terms of adsorption strength (displaying effects of bridging between different branches in denser areas), whereas the latter is better in terms of adsorption capacity. Our plausible answer for the question why the hyperbranched architectures are found to be more efficient is that because they contain both star-like and comb-like local fragments, and, therefore, may benefit of combining strong sides of both types of branching.

The outlook of this study may be sketched as examination of the other types of the core-shell architectures with higher dimensionality and special symmetry of a core; as well as the other types of branching, e.g. dendritic or bottle-brush structures, in search for optimal adsorption effectiveness.

\section*{Acknowledgments}
The work was supported by Academy of Finland, reference number 334244, and by the NRFU Grant No. 2023.05/0019.

{\bf{Data availability statements}}
The data that support the findings of this article are openly available  \cite{reposit}

\bibliographystyle{apsrev-nourl}
\bibliography{Chelation_branched}
\end{document}